\documentclass{article}

\usepackage{arxiv}

\usepackage[utf8]{inputenc} % allow utf-8 input
\usepackage[T1]{fontenc}    % use 8-bit T1 fonts
\usepackage{hyperref}       % hyperlinks
\usepackage{url}            % simple URL typesetting
\usepackage{booktabs}       % professional-quality tables
\usepackage{amsfonts}       % blackboard math symbols
\usepackage{nicefrac}       % compact symbols for 1/2, etc.
\usepackage{microtype}      % microtypography
\usepackage{lipsum}
\usepackage{graphicx}
\graphicspath{./Figure/}
\usepackage{subcaption}
\usepackage{ragged2e}
\usepackage{tabularx}
\usepackage{amsmath}
\usepackage{mathtools}
\usepackage{cuted}
\usepackage{amssymb}
\usepackage{lipsum}
\usepackage{nccmath}
\usepackage{autobreak}
\usepackage{makecell}
\usepackage{xurl}
\usepackage{xcolor}

 \title{Exploring Crowdworkers' Perceptions, Current Practices, and Desired Practices Regarding Using Non-Workstation Devices for Crowdwork}

\author{
 Senjuti Dutta \\
  University of Tennessee, Knoxville\\
  \texttt{sdutta6@vols.utk.edu} \\
   \And
Scott Ruoti \\
   University of Tennessee, Knoxville\\ 
     \texttt{ruoti@utk.edu}  \\
    \And
 Rhema Linder \\
University of Tennessee, Knoxville\\ 
  \texttt{rlinder@utk.edu} \\
  \And
 Alex C. Williams \\
  Amazon AI \\
  \texttt{acwio@amazon.com} \\
 \And
Anastasia Kuzminykh \\
    University of Toronto\\
    \texttt{anastasia.kuzminykh@utoronto.ca} \\
}
\begin{document}

\maketitle

\begin{abstract}
%Crowdsourcing platforms predominantly focus on workstation interfaces, limiting the flexibility of crowdworkers. Recognizing the need for more adaptable platforms, previous work has identified key characteristics of work processes that vary among crowdworkers, including device type and stage of work. However, these variables have typically been explored independently. Our study is the first to examine their interrelated variabilities among crowdworkers.
%We utilized a survey involving 150 Amazon Mechanical Turk crowdworkers to obtain insights, revealing three distinct groups characterized by interrelated variabilities in these key aspects.
%The largest group of workers' work practices imply a dependence on traditional devices with limited openness to use smartphone and tablet.
%The second-largest group of workers' work practices also imply traditional device usage, however prefer support for tools and scripts across all devices, especially for smartphone and tablet.
%The small group of workers  currenly use as wel as have strong preference in using non-workstation devices, especially smartphone and tablet.
%We contextualize our findings into design insights for platform designers, discussing the corresponding implications for creating more personalized, flexible, and efficient crowdsourcing platforms. Additionally, we present the unique work practices of these clusters of crowdworkers  in contrast to information workers.

Despite a plethora of research dedicated to designing HITs for
non-workstations, there is a lack of research looking
specifically into workers’ perceptions of the suitability of
these devices for managing and completing work. In this
work, we fill this research gap by conducting an online survey
of 148 workers on Amazon Mechanical Turk to explore (i)
how crowdworkers currently use their non-workstation
devices to complete and manage crowdwork, (ii) what
challenges they face using those devices, and (iii) to what
extent they wish they could use those devices if their concerns
were addressed. Our results show that workers unanimously
favor using a desktop to complete and manage crowdwork.
While workers occasionally use smartphones or tablets, they
find their suitability marginal at best and have little interest in
smart speakers and smartwatches, viewing them as unsuitable
for crowdwork. When investigating the reason for these views,
we find that the key issue is that non-workstation devices lack
the tooling necessary to automatically find and accept HITs,
tooling that workers view as essential in their efforts to
compete with bots in accepting high-paying work. To address
this problem, we propose a new paradigm for finding,
accepting, and completing crowdwork that puts crowdworkers
on equal footing with bots in these tasks. We also describe
future research directions for tailoring HITs to
non-workstation devices and definitely answering whether
smart speakers and smartwatches have a place in crowdwork.
\end{abstract}
\section{Introduction}

Crowdwork consists of various tasks performed by individuals, all working towards a shared objective.
These tasks, often called Human Intelligence Tasks (HITs), are published by requesters on crowdsourcing marketplaces to find crowdworkers who will complete the task in exchange for a monetary reward.
Amazon's Mechanical Turk (MTurk) is one such marketplace.
MTurk was originally focused on helping AI researchers build training data sets by having many crowdworkers label data.
While MTurk is still used for this purpose, it has evolved to be used by researchers in many other disciplines that need to collect data from many participants, such as conducting surveys or executing usability studies.

People are drawn to crowdwork for varied reasons, such as contributing to scientific advancement or combatting boredom.
However, it has increasingly become a principal source of income for many workers~\cite{posch2019measuring}.
Within this context, it is no surprise that workers frequently experience frustration and challenges related to crowdwork, such as demanding time pressure for task completion~\cite{yin2018running}.
As researchers directly benefit from the results of crowdwork, we believe the community has an ethical responsibility to understand and address the challenges facing crowdworkers.

One key area for enhancement is worker flexibility, allowing them to manage and complete work at the time, location, and manner of their choosing.
Prior research into flexibility in traditional work environments has demonstrated that it positively impacts working conditions, productivity, and job satisfaction~\cite{baltes1999flexible,neirotti2019designing}.
We expect crowdworkers would reap similar benefits if their flexibility could be improved.

Previous research~\cite{dutta2022mobilizing,hettiachchi2020context} has suggested that non-workstation devices could strongly influence worker flexibility. 
Despite a plethora of research dedicated to designing HITs for non-workstation devices~\cite{eagle2009txteagle,narula2011mobileworks,hettiachchi2020hi,nebeling2016wearwrite}, there is a lack of research looking specifically into workers' perceptions of the suitability of these devices for managing and completing work.
Motivated by this unaddressed gap, we delve deeply into the nuanced perspectives of crowdworkers, exploring the efficacy and challenges of non-workstation devices in promoting or possibly curtailing their work adaptability.

To investigate this question, we surveyed 148 MTurk crowdworkers regarding their perceptions of managing and completing crowdwork using workstations, smartphones, tablets, smart speakers, and smartwatches.
We employ a mixed-methods approach, combining quantitative and qualitative research techniques.
The survey questions explore (i) how crowdworkers currently use their non-workstation devices to complete and manage crowdwork, (ii) what challenges they face using those devices, and (iii) to what extent they wish they could use those devices if their concerns were addressed.
Our key findings include,

\begin{enumerate}
	\item \textbf{Crowdwork is workstation-centric, with no other devices suitable to replace the workstation, even temporarily.}
	Nearly all workers rely primarily on their workstations to complete crowdwork, with workers finding workstations nearly perfectly suited to these tasks.
	In contrast, while up to a quarter of participants use smartphones to complete or manage work occasionally, they find the suitability of these devices to be marginal at best.
	In fact, if forced to use these devices due to a malfunctioning workstation, two-thirds of workers would rather stop working.
	
	\item \textbf{Competing with bots for high-paying work is central to workers' unwillingness to use non-workstation devices.}
	During our thematic analysis of workers' qualitative answers regarding the challenges of using non-workstation devices, it became clear that outracing bots to high-paying work was critical context for their other challenges.
	Workers' solution to this problem was using tooling that automatically finds and accepts crowdwork.
	Workers were unwilling to use non-workstation devices for crowdwork management because these tools are only available on workstations.
	Moreover, due to tight timelines for completing work after it is auto-accepted, workers felt the need to complete work on the device used to manage work---i.e., the workstation.
	
	\item \textbf{There is a need for improved HIT designs that support non-workstation devices.}
	While workers are hesitant to complete work on a device other than the one where they auto-accept work, when they do complete work on a non-workstation device, they wish it was better formatted for those devices.
	The most common concern was that the HITs were non-responsive, failing to display elegantly on a smartphone or tablet screen or, in the (frequent) worst case, being entirely inoperable on those devices.
	The other major concern was that completing surveys (the most common non-workstation HIT) using virtual keyboards was difficult.
	
	\item \textbf{Workers are not interested in using smart speakers or smartwatches to complete work.}
	Participants labeled smart speakers and smartwatches as highly unsuitable for completing or managing crowdwork.
	Half of the participants even indicated that in a world where crowdwork was tailored to work on these devices, they still couldn't conceive of wanting to use those devices to complete work.
	This starkly contrasts prior work~\cite{nebeling2016wearwrite,hettiachchi2020hi,hettiachchi2020context} that showed workers being open to using smart speakers for crowdwork.
\end{enumerate}

Based on these findings, we conclude the paper by discussing how challenges around using non-workstation devices could begin to be ameliorated.
We also discuss potential areas of future research based on our results.

\section{Related Work}

While the vast majority of crowdwork is performed using workstations~\cite{dutta2022mobilizing}, prior work has explored the feasibility of using non-workstation devices for completing crowdwork.

\begin{itemize}
	\item TxtEagle~\cite{eagle2009txteagle} and mClerk~\cite{gupta2012mclerk} are two platforms that support completing tasks using SMS text messages.
	
	\item Yan et al.introduced an iOS application supporting sensor-related crowdsourcing tasks using smartphones~\cite{yan2009mcrowd} . 
	
	\item MobileWorks~\cite{narula2011mobileworks} is a smartphone-based crowdsourcing platform that administers optical character recognition (OCR) tasks. Later Kumar et al.introduced a mobile-based crowdsourcing platform designed to work on Android devices for different categories of crowdsourcing tasks (e.g., human OCR, image tagging, language translation)~\cite{kumar2014wallah}.
	
	\item Respeak~\cite{vashistha2017respeak} and Bspeak~\cite{vashistha2018bspeak} are voice-centric systems aimed at transcribing audio files using a smartphone. %Bspeak emphasizes converting files specifically through verbal speech.
	
	\item Hettiachchi et al.demonstrated that smart speakers could be used to complete some crowdwork tasks and have important flexibility benefits\cite{hettiachchi2020hi} .
	
	\item Nebeling et al. built a crowdwork system for collaborating text using voice dictation on smartwatches~\cite{nebeling2016wearwrite} .
\end{itemize}

While the above work has demonstrated the feasibility of executing crowdsourcing tasks on non-workstation devices, it did not measure crowdworkers interest in using non-workstation devices for completing crowdwork.
Exploring this question, Dutta et al. surveyed 104 MTurk workers, finding that most were interested in using smartphones for completing crowdwork\cite{dutta2022mobilizing} .
Similarly, Hettiachchi et al. surveyed 329 MTurk workers, finding that roughly three-quarters of participants were open to using smartphones~\cite{hettiachchi2020context} and smart speakers for completing crowdwork.
Conversely, previous research found a general reluctance to use non-workstation devices for completing crowdwork~\cite{williams2019perpetual,newlands2021crowdwork} .
\emph{In our work, we extend prior work by not only considering current perceptions and practices but also measuring how interested participants would be in using non-workstation devices for completing crowdwork in a world where crowdwork has been tailored to those devices.}

%\subsection{HIT Management on Non-Workstation Devices}

Non-workstation devices can also be used to manage crowdwork.
In interviews with 21 MTurk workers, Williams et al. found that many of these workers already use smartphones and tablets to manage work~\cite{williams2019perpetual} .
Similarly, Newlands et al. \cite{newlands2021crowdwork} surveyed 605 MTurk workers from the US and India, finding that many used smartphones and tablets to manage their work.
However, in both studies, task management on non-workstations was an ancillary result, and user practices were not explored in depth.
\emph{In our work, we examine this issue in-depth, measure current and desired practices, and identify pain points related to using non-workstation devices to manage work.}
\section{Methodology}

We conducted a survey targeting MTurk workers, aiming to uncover the following:
(i) their current and desired practices for using workstations, smartphones, tablets, smart speakers, and smartwatches to manage and complete tasks,
(ii) the challenges they face while working on these devices, and
(iii) their suggestions for improving these devices to enhance their work flexibility.
This survey was conducted on November 9, 2021, and participants were compensated USD \$5.00 for their participation.
We collected 150 responses in total, and after removing two for data quality issues, we were left with 148 valid responses. 
% \textcolor{purple}{We ensured that the survey task was accessible to crowd workers on any device, including desktops, smartphones, and tablets. No device restrictions were applied in the task configuration on MTurk, allowing participants to complete the survey using whichever device they typically use for crowd work. This approach was taken to minimize the risk of a biased study sample and to capture a diverse range of device usage among participants.}

% The IRB-approved survey is available as supplemental material.
% We will publish our data and analysis scripts along with this paper.

\subsection{Survey Content}

In Section 1, we collected participants' demographic details, including age, gender, and educational background. 
We asked about their experience on MTurk, including their number of completed HITs, hours worked a week, and HIT approval rates.
This section concluded with questions about their specific tools for managing and executing crowdwork.

In Section 2, we asked participants how often they use a workstation, smartphone, tablet, smart speaker, or smartwatch to complete crowdwork.
We also investigate the impact of location on crowdwork completion~\cite{hettiachchi2020context} by having participants indicate how often they use these devices to complete crowdwork when at their workstation, away from it, or away from home.
Next, we asked participants to state the types of crowdwork they complete using these devices (open response).
This was followed by having them rate (on a scale of 1--5) the usability of completing traditional AI-training-based tasks~\cite{hettiachchi2020context}: sentiment analysis, information finding, audio tagging, speech transcription, image classification, bounding box.
Furthermore, we asked what types of crowdwork they wished were better optimized for these devices (open response).
We concluded this section by asking them how they choose which device to use to complete crowdwork (open response).

In Section 3, participants indicated how frequently they used the aforementioned devices for crowdwork management.
They also rated (on a scale of 1--5) the usability of completing crowdwork management tasks~\cite{williams2019perpetual} on these devices: finding crowdwork, auto-accepting crowdwork, creating catchers/watchers, listening to catchers/watchers, talking with other MTurk workers, talking to requesters.
We also asked them to indicate which tasks they would like better supported.
%We finished by asking them how they choose which device to use to manage their HITs.

In Section 4, we inquired about the inconvenience participants would experience if their primary workstation was rendered non-functional.
We then asked which alternative device they would use in this situation and why (open response).
Next, participants shared how their workstation and alternative device compared in terms of acceptability and effectiveness.
We also asked how easy it was to transfer from the workstation to the alternative device.

In Section 5, the final section, we posed a hypothetical scenario to participants.
For each device, we asked them how they would reimagine or optimize any facet of crowdwork if equipped with a magical tool to do so (open response).

\subsubsection{Survey Development}

After creating an initial version of our survey, we improved it over several rounds of internal review from our team.
Once we produced a version we were satisfied with, we submitted for and received IRB approval for our study.
We piloted the survey with a group of 20 individuals.
Finding no substantial issues, we proceeded to launch the finalized survey.

\subsubsection{Quality Control}
During the qualitative data coding process, we identified two respondents who provided single-word responses to all open-ended questions, which lacked coherence. Consequently, we excluded these two participants, resulting in a final dataset comprising 148 participants for our analysis.

% \textcolor{purple}{We initially excluded two participants who provided incoherent single-word responses. To further ensure data quality, we carefully reviewed the remaining responses for coherence and consistency. This process led to the exclusion of additional participants, resulting in a final dataset of 148 participants, thereby enhancing the study's robustness.}

\subsection{Demographics}

%\input{Tables/demographics}

%Table \ref{tab:demographics} shows the breakdown of the demographics of our survey, including the demographics overall and based on what devices participants owned.
Our participant pool was two-thirds male (65\%).
Participants skewed younger: 18--24 (1\%), 25--34 (43\%), 35--44 (28\%) and 45--54 (18\%), 55--64 (8\%), 65+ (1\%).
Half had a bachelor's degree: high school (27\%), some college (16\%), bachelor's degree (51\%), or advanced degree (7\%).
Most had been on MTurk for 2--5 years: Less than 1 (7\%), 1--2 (14\%), 2--5 (49\%), 5+ (30\%).
Finally, on average, participants had twenty-five thousand completed HITs with a median 100\% completion rate.

Looking at the devices they owned, nearly all had a workstation ($n=147$;~99\%), and the majority had a smartphone ($n=145$;~98\%), a tablet ($n=125$;~84\%), a smart speaker ($n=121$;~82\%), or a smartwatch ($n=109$;~74\%).
% There were no meaningful differences in demographics based on what device the participants owned.
% \textcolor{purple}{The distribution of gender, age, educational attainment, MTurk experience, and weekly hours spent on MTurk was consistent across different device ownership categories, indicating that demographic factors did not meaningfully vary by the type of device participants used}.

\subsection{Data Analysis}
Qualitative data was analyzed using thematic analysis~\cite{boyatzis1998transforming}.
Two researchers collaboratively conducted initial coding, systematically identifying level-one codes based on participants' textual responses~\cite{gioia2013seeking}.
Second, the researchers grouped similar codes into higher-level concepts.
Finally, the researchers identified themes by connecting these concepts and delineating the overarching themes discussed in our findings.
Throughout the process, the researchers took detailed coding notes and memos.

Whenever we report the percentage of participants that said something about a particular device, we base this percentage on the number of participants that owned that device.
When quoting from participants, we label them as P1--150, based on their position in the data prior to removing bad data.

\subsection{Limitations}

Our work is exploratory, only surveying 150 participants and having a significant qualitative component.
As with all such work, future work is needed to quantify these results at a scale sufficient to measure the generalizability of our findings.

Our research is targeted at Amazon MTurk crowdworkers.
Future research is needed to explore to what extent our findings apply to crowdworkers on other platforms.

Finally, our results are based on self-reported data, which may not precisely measure how often participants used a particular device. However, we believe it is likely accurate enough for our exploratory study.
Similarly, our qualitative data includes participants' suppositions about how they would like to use devices in an ideal crowdworking world.
While such data does not indicate how users would act in reality, it provides insights into users' thinking and desires, which is what we were seeking in our work.

\section{Results}

%In this section, we report on the quantitative and qualitative results from our survey of 148 Amazon Mechanical Turk (MTurk) workers.
%These results are organized based on the order we asked questions.

In this section, we first cover the current usage of devices to complete and manage crowdwork.
We then examine impediments that prevent increased usage of non-workstations to complete and manage crowdwork.

\subsection{Current Usage}

\begin{figure}
	\centering
	\includegraphics[width=\columnwidth]{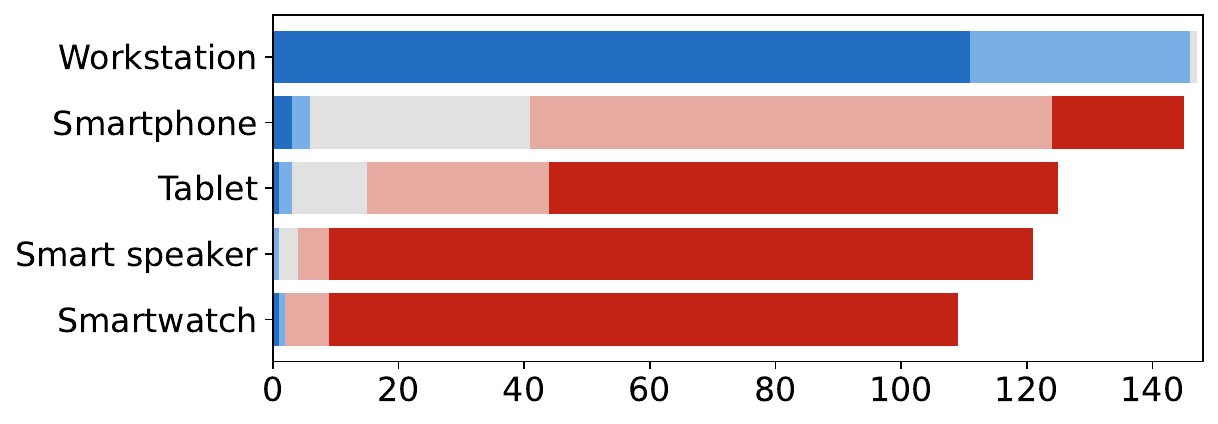}
	\includegraphics[width=\columnwidth]{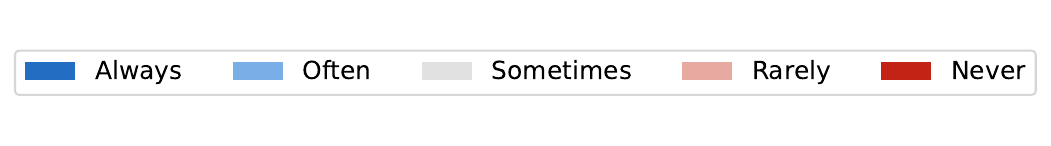}
	\caption{Device usage for completing crowdwork}
	\label{fig:device_usage}
\end{figure}

\begin{figure}
	\centering
	\includegraphics[width=\columnwidth]{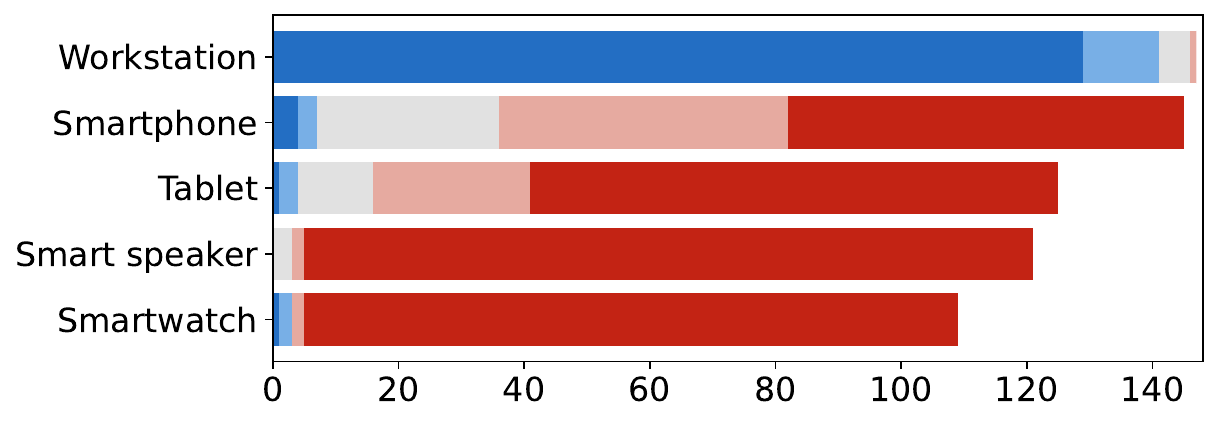}
	\includegraphics[width=\columnwidth]{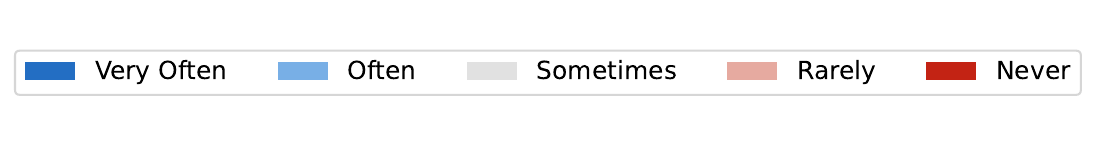}
	\caption{Device usage for managing crowdwork}
	\label{fig:device_usage_task_management}
\end{figure}

While a majority of our participants owned all five types of devices (workstation, smartphone, tablet, smart speaker, smartwatch), workstations were where participants reported completing (99\%) or managing (96\%) work `always' or `often' (see Figures~\ref{fig:device_usage}~and~\ref{fig:device_usage_task_management}, respectively).
Still, roughly a quarter of participants who owned a smartphone reported using it at least occasionally to complete (28\%) or manage (25\%) crowdwork.
Similarly, a tenth of participants who owned tablets reported using it to occasionally complete (12\%) or manage (13\%) crowdwork.
Smart speaker and smartwatch usage was nearly non-existent.
% \textcolor{purple}{The small percentage of participants who did mention using these devices primarily for completing tasks did so when the task specifically required it, as indicated by their responses to questions about the types of tasks they currently perform using speakers and watches in Figure ~\ref{fig:current_hit_types}}.

\begin{figure}[t]
	\centering
	\includegraphics[width=\columnwidth]{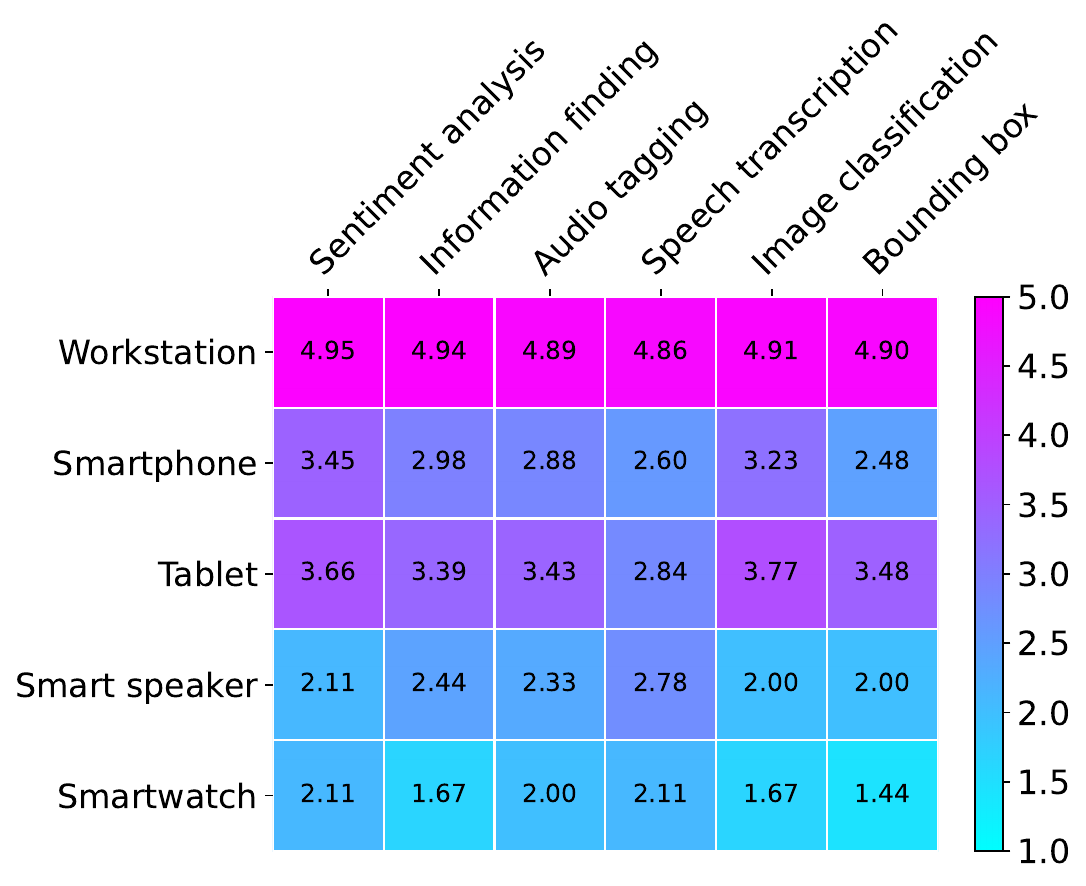}
	\caption{Crowdwork suitability by device}
	\label{fig:task_suitability}
\end{figure}

\begin{figure}[t]
	\centering
	\includegraphics[width=\columnwidth]{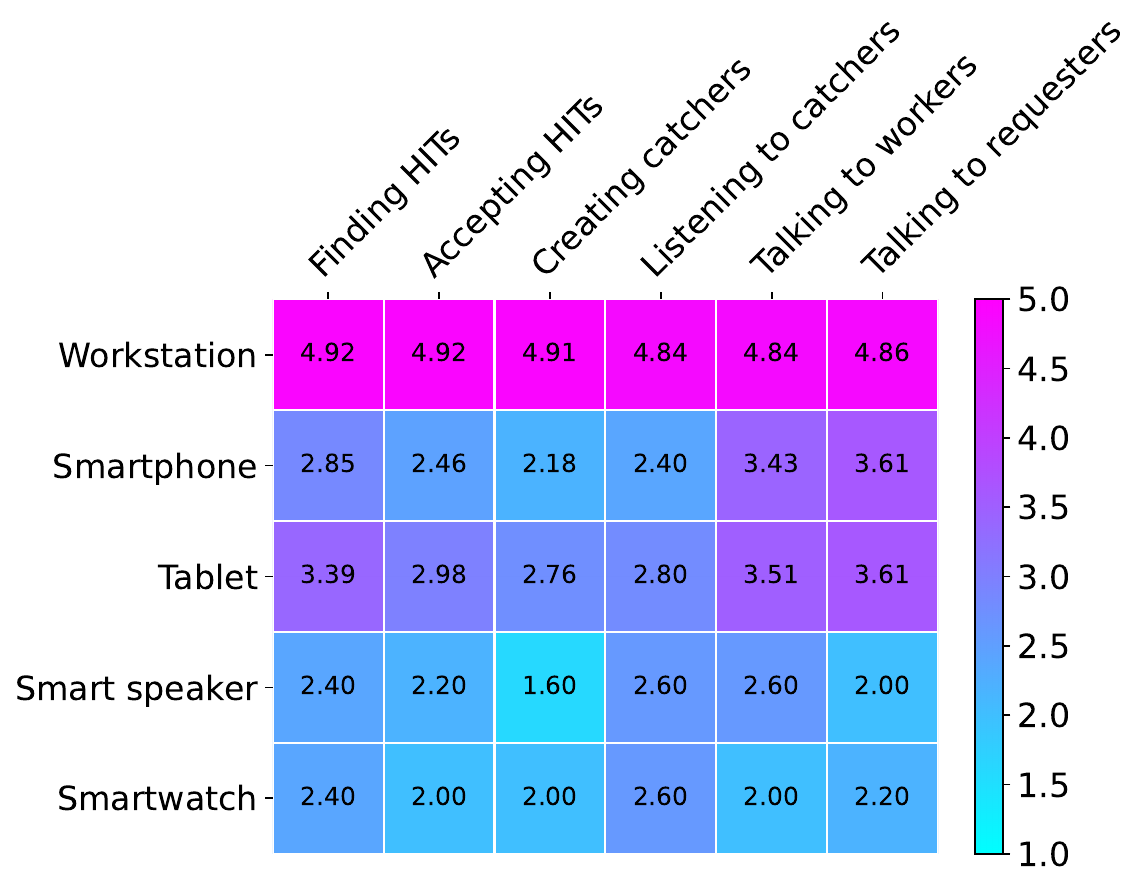}
	\caption{Crowdwork suitability by device}
	\label{fig:task_management_effectiveness}
\end{figure}

These usage patterns closely align to how suitable participants found each device for completing (see Figure~\ref{fig:task_suitability}) or managing (see Figure~\ref{fig:task_management_effectiveness}) crowdwork.
In both cases, workstations were considered as highly suitable for completing and managing work (4.9/5).
In contrast, perceptions of tablets skewed slightly positive for completing (3.4/5) and managing (3.2/5) crowdwork.
Smartphones skewed slightly negatively for completion (2.9/5) and management (2.8/5).
Smart speakers (2.3/5 and 2.2/5) and smartwatches (1.8/5 and 2.2/5) were viewed negatively.

\begin{figure}
	\centering
	\includegraphics[width=\columnwidth]{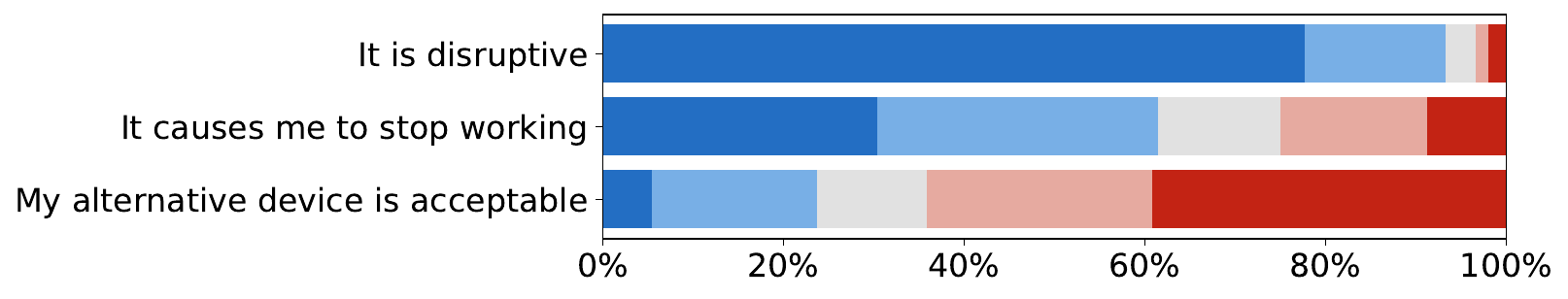}
	\includegraphics[width=\columnwidth]{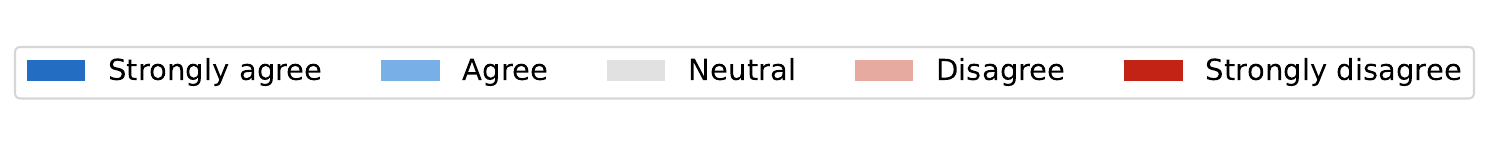}
	\caption{Reaction to workstation being unavailable}
	\label{fig:broken_workstation}
\end{figure}

%\begin{figure}
%	\centering
%	\includegraphics[width=.4\textwidth]{broken_worsktation_selection}
%	\caption{How participants select which device they will use if their workstation is broken.}
%	\label{fig:broken_worsktation_selection}
%\end{figure}

To understand the impact on crowd workers when using non-workstation devices, we investigated what workers do when their workstation becomes unavailable.
In this event, participants said they would switch to using a smartphone (53\%) or a tablet (47\%).
However, nearly all participants (98\%) indicated that this would be highly disruptive to their ability to complete and manage work, with two-thirds (61\%) saying they would simply stop doing crowdwork until their workstation was available again (see Figure~\ref{fig:broken_workstation}).
Such answers make sense, as only a quarter of participants (24\%) find their alternative device an acceptable substitute.

\subsubsection{Non-Workstation HIT Types}

\begin{figure}
	\centering
	\includegraphics[width=\columnwidth]{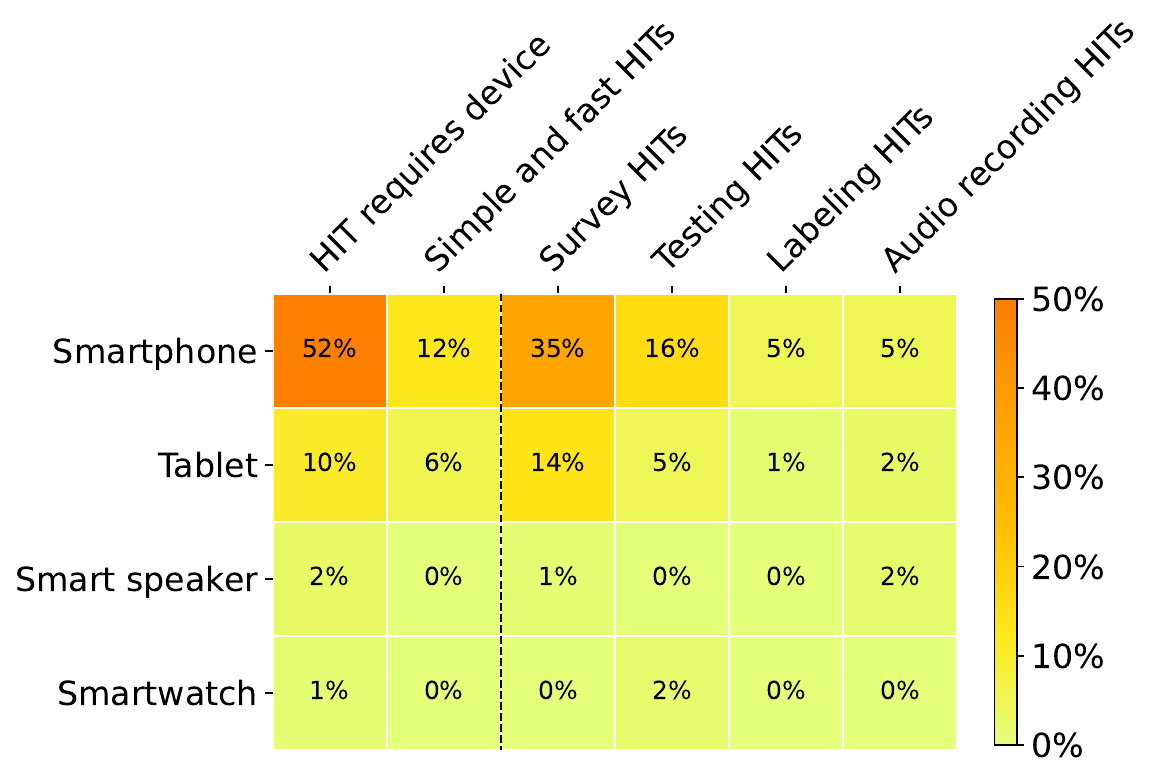}
	\caption{HIT types completed on smartphone and tablet (open response question, coded in thematic analysis)}
	\label{fig:current_hit_types}
\end{figure}

When asking participants what type of HITs they complete on non-workstation devices (see Figure~\ref{fig:current_hit_types}), many indicated that they would only do so if the requester specifically required it (52\%~smartphone, 10\%~tablet):

%\begin{quote}
%	\textit {``Mostly whether [the smartphone is] required or not by the requester. Otherwise, I only really use my phone if the HIT requires audio/video collection or uploads, as this is usually easier on mobile.''} (P118)
%\end{quote}

\begin{quote}
	\textit {``Only [if a smartphone] it is absolutely required. A full keyboard is needed in most cases.''} (P114)
\end{quote}

\begin{quote}
	\textit {``When I'm required to complete them on a smartphone. Most are testability type HITs for mobile web pages, but for a few I will download an App for testing, but only if I really trust the source of the App.''} (P82)
\end{quote}

The most common type of HIT was surveys (35\%~smartphone, 14\%~tablet), followed by application testing (16\%~smartphone, 5\%~tablet).
In general, participants focused on HITs that were simple and fast to complete (12\%~smartphone, 6\%~tablet).
Interestingly, participants largely omitted mention of AI-training-related HITs.

\subsection{Impediments to Adopting Non-Workstation Devices}

Our quantitative data shows that users find non-workstation devices unsuitable for crowdwork and largely avoid using them.
Next, we turn to our qualitative data and thematic analysis to ascertain what exactly is problematic with these devices and what would need to change to allow workers to integrate these devices into their workflows.

Surprisingly, the central theme that emerged from our analysis was not about the devices themselves but rather an ecosystem-wide issue---finding enough work.
More specifically, workers struggle to outrace bots to claim high-paying work.
We start our discussion of the impediments to adopting non-workstation devices with this topic, as it provides crucial context for understanding participants' challenges and desires regarding non-workstation devices.

\subsubsection{Competing with Bots}

Participant responses frequently mentioned struggling to find enough work.
Bots were commonly mentioned in regard to this issue: 

\begin{quote}
	\textit{``I would take the bots away so the real people have a better chance at earning money. It gets really old when HITs are taken in an instant constantly.} (P37)
\end{quote}

\begin{quote}
	\textit{``Workers are frustrated by the lack of work provided, especially when considering bots take some of the work.''} (P62)
\end{quote}

%\begin{figure}
%	\centering
%	\begin{minipage}{.45\textwidth}
	%		\centering
	%		\includegraphics[width=\textwidth]{tools_per_participant}
	%		\captionof{figure}{The number of tools each participant uses.}
	%		\label{fig:tools_per_participant}
	%	\end{minipage}\hfill%
%	\begin{minipage}{.45\textwidth}
	%		\centering
	%		\includegraphics[width=\textwidth]{tool_counts}
	%		\captionof{figure}{The most common tools used by our participants.}
	%		\label{fig:tool_counts}
	%	\end{minipage}
%\end{figure}

Workers primary response to this problem was to leverage tooling that allowed them to automate the process of finding and accepting crowdwork.
The vast majority of workers used at least one tool for this purpose, with nearly half using two or more: no tools (14\%), 1 tool (40\%). 2 tools (28\%), 3 tools (11\%), 4+ tools (7\%).
The most common tools were MTurk Suite (64\% of participants), Panda Crazy (25\%), HIT Forker (14\%), Queuebicle (12\%), and TurkerView (11\%).

However, the practice of automatically catching and accepting work has its challenges.
After the tooling accepts work, workers must complete that work within a requester-selected time frame.
This was a common frustration for workers and often left them feeling like they were tied to their workstations when using auto-catcher tools:

\begin{quote}
	\textit{``Honestly, my only change would probably have to be the timers a requester gives their HIT. if I have to leave my home and end up catching a good HIT while I'm gone, I almost always miss it. I want longer timers.''} (P63)
\end{quote}

\begin{quote}
	\textit{``I would have requestors optimally recommend for more time that someone can keep a HIT which will allow people to better manage them.''} (P25)
\end{quote}

%\begin{quote}
%	\textit{``I would give workers more time, guarantee individual work based on qualifications. The reasons are because some requesters do not provide enough time to finish HITs, so workers feel rushed.''} (P62)
%\end{quote}

\subsubsection{Lack of Tooling on Non-Workstation Devices}

%\begin{figure}
%	\centering
%	\includegraphics[width=\columnwidth]{magic_wand}
%	\caption{Percent of participants requesting improved support for crowdwork on non-workstation devices}
%	\label{fig:magic_wand}
%\end{figure}

With the above context in mind, it came as no surprise that when asked what they would change about crowdwork on non-workstation devices, a third of participants indicated wanting crowdwork tools to be available on smartphones (37\%) and tablets (33\%).
As these codes are derived from unguided, open-response questions, the number of users who would actually desire this is likely higher (i.e., these numbers are lower bounds).

\begin{figure}
	\centering
	\includegraphics[width=\columnwidth]{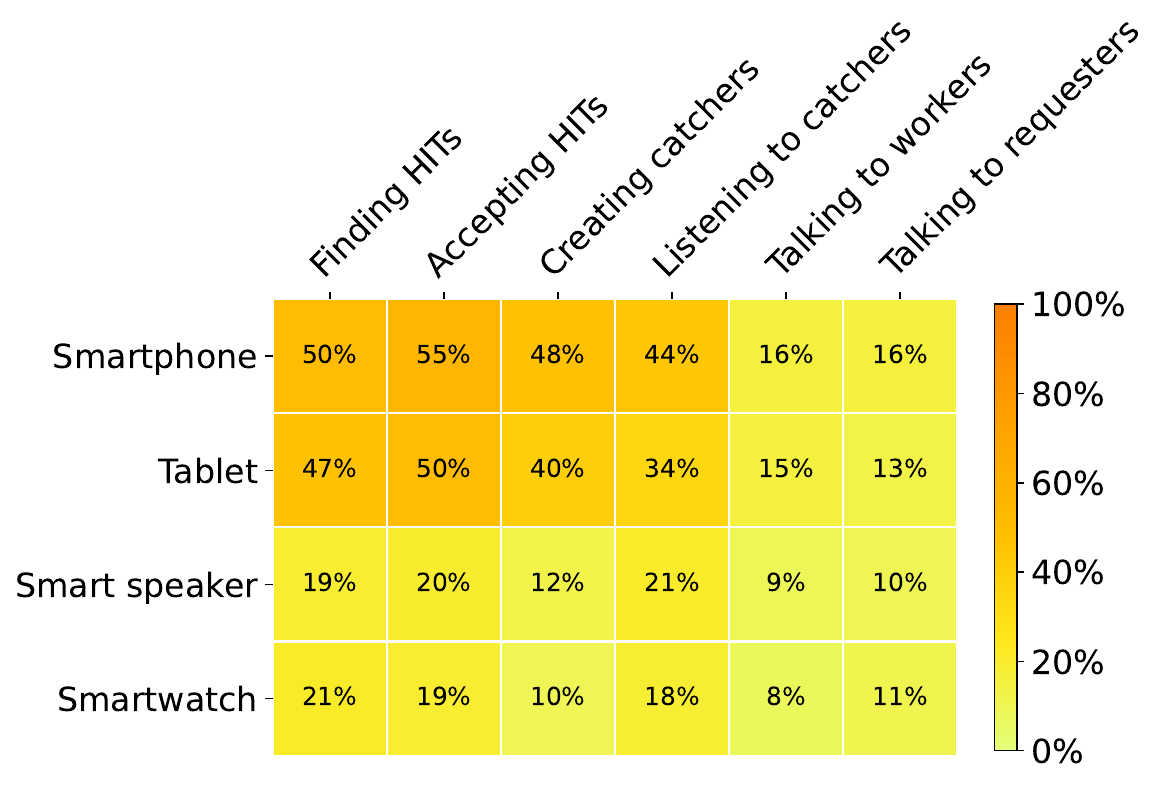}
	\caption{Percentage of participants who desire improved support for crowdwork management on that device}
	\label{fig:task_management_wants_support}
\end{figure}

When asked specifically what work management tasks need to be better supported on non-workstation devices, half of participants wanted better support for automatically filtering and accepting HITs on smartphone and tablets (see Figure~\ref{fig:task_management_wants_support}).
This was supported in the qualitative data where auto-accepting HITS (12\%~smartphone, 12\%~tablet) and HIT filtering (7\%~smartphone, 8\%~tablet) were common topics: 

\begin{quote}
	\textit {``I would add HIT catchers and scripts to the smartphone platform. If MTurk suite can run in Chrome on a PC, it should be able to run on Chrome on a smartphone.''} (P50)
\end{quote}

\begin{quote}
	\textit {``[For my tablet, the] same as my smartphone: scripts, ratings, catchers, etc.''} (P51)
\end{quote}

In regards to finding HITs, there was a desire for improved tooling that could filter for HITs that paid a good hourly wage and were offered by requesters that were unlikely to reject the HIT:

\begin{quote}
	\textit{``I would have HITs rated on \$ per hour. I currently auto-catch anything that is above 50c, however sometimes I see that these hits are 50c for something like 30 mins, which I find ridiculous. The minimum rate I am usually willing to work is 10c/minute. If I could easily catch any HIT that is over 10c/minute, I think that would help me out a lot.''} (P65)
\end{quote}

\begin{quote}
	\textit{``I would just improve the existing tools I have, especially Turk Guru. Sometimes it accepts HITs from requesters who have acceptance ratings below my minimum threshold or don't pay my minimum hourly rate. That aspect could be improved.'' (P68)}
\end{quote}

%\begin{quote}
%	\textit{``Have better access to requester warnings (non payment, rejection) and accurate pay rates.'' (P114)}
%\end{quote}

For smart speakers and smartwatches, roughly one-fifth of participants also wanted better support for automatically filtering and accepting HITs.
Our qualitative data suggests that most of this interest centered around being notified of HITs matching their filtering criteria and being able to accept those HITs on their smart speaker (8\%) or smartwatch (6\%):

\begin{quote}
	\textit{``\dots Workers may be away from their workstations or house, but can carry the smartwatch with them, allowing them to catch HITs \dots.''} (P62)
\end{quote}

\begin{quote}
	\textit {``Setting up alerts and listening/getting a notification for them, being able to speak to accept HITs [on my smart speaker] when I'm away.''} (P83)
\end{quote}

%\begin{quote}
%	\textit{``I guess if [a smart speaker] could just notify me if a requester I like has posted work that would be nice.''} (P128)
%\end{quote}

\begin{quote}
	\textit{``I'd love to be able to hear from my smart speaker anytime a HIT that matches my preferences is available, then ask whether I'd like them to accept the HIT or not.''} (P12)
\end{quote}

%\begin{quote}
%	\textit{``I would allow for notifications from Mturk to go through to your watch to alert you of expiring hits or new hits accepted.''} (P25)
%\end{quote}

%
%Participants also want better support for HIT filtering (24\%):
%
%\begin{quote}
%	\textit {`` I would make it easier to filter out poor paying or low-quality requesters, because I spend a lot of time filtering out junk''} (P50)
%\end{quote}
%
%
%\begin{quote}
%	\textit {``Better filtering choices to select preferred types of hits.''} (P110)
%\end{quote}
%
%\begin{quote}
%	\textit {``Better and easier native ways to categorize and filter HITs and flag or rate requesters.''} (P119)
%\end{quote}

\subsubsection{Need for Improved HIT Design}

While our thematic analysis identified the need for tooling on non-workstation devices to compete with bots for high-paying work as the primary concern for workers, a close second was a need for crowdwork to be better designed to support non-workstation devices.
Most of this focused on improving HIT design for smartphones (35\%) and tablets (26\%).

Participants gave multiple reasons why using a smartphone or tablet is problematic.
One common reason (18\%) was that these devices' screens were simply too small, often leading to problems displaying the HIT (19\%).
This is backed up by the fact that when participants need to select an alternative device to use when their workstation is broken, their primary concern is the alternative device's screen size (32\%):

\begin{quote}
	\textit {``It is much easier to do HITs and browse Mturk on my desktop. Phone/tablet screens are too small to get any work done.''} (P142)
\end{quote}

\begin{quote}
	\textit {``Depending on the screen size and supported features on the device. Most of those devices are not even compatible for such hits. I always prefer to use my desktop or laptop due to the size and compatibility.''} (P146)
\end{quote}

\begin{quote}
	\textit{``I would format HITs to fit the screen [on my tablet] \dots This would make it easier to complete tasks and find tasks.''} (P87)
\end{quote}

%\begin{quote}
%	\textit{``I would like [my smartphone] to be easier to accept HITs and to be able to actually do them. I would like for it to be formatted to my phone so I can navigate easier.''} (P14)
%\end{quote}

Another common issue (18\%) was that participants found entering text on a smartphone or tablet to be difficult.
This was especially pronounced when completing surveys, which often require significant textual entry (as our survey did):

\begin{quote}
	\textit {``A full keyboard is needed in most cases.''} (P114)
\end{quote}

A full keyboard is needed in most cases.

\begin{quote}
	\textit {``I also prefer a real keyboard and larger screen. It takes way too long to complete the same amount of work on anything else.''} (P64)
\end{quote}

\begin{quote}
	\textit {``Anything I have to type on I'm using a computer if at all possible since I have large fingers, and am 68 and didn't grow up texting on a phone.''} (P82)
\end{quote}

\subsubsection{Smart Speakers and Smartwatches Skepticism}

%\begin{figure}
%	\centering
%	\includegraphics[width=\columnwidth]{desired_hit_types}
%	\caption{Types of HITs participants are interested in completing with different devices. Data is from an open-response question. Percentages are based on the number of participants that have used each device to complete HITs.}
%	\label{fig:desired_hit_types}
%\end{figure}

Nearly half of our participants indicated that they had no interest or thought it was infeasible to use smart speakers (45\%) or smartwatches (50\%) to complete crowdwork.
While roughly a quarter of participants had similar sentiments regarding smartphones (20\%) and tablets (27\%), our thematic analysis found that this sentiment was much more pronounced and negative for smart speakers and smartwatches:

\begin{quote}
	\textit {``Same as with the speaker, I really don't see how this would work. The watch interface is a pain in the behind just for being a watch, I can't imagine trying to fill out surveys with it. \ldots Nothing is going to improve this on a smart watch aside from making the watch face the size of a tablet.''} (P5)
\end{quote}

\begin{quote}
	\textit {``I cannot think of a way to do this because I believe its stupid to use this platform on a smart speaker.''} (P103)
\end{quote}

\begin{quote}
	\textit {``I don't think anything would work out on a smartwatch.''} (P76)
\end{quote}

This negative sentiment, combined with the fact that we find little interest in using smart speakers or smartwatches, even in a world where crowdworkers had a magic wand to fix the issues with completing or managing crowdwork on these device, suggests that getting users to use these devices for crowdwork is likely an uphill battle.
Still, a small but non-negligible number of our participants wanted HITs to be designed to better support smart speakers (7\%) or smart watches (8\%).
This indicates that a body of crowdworkers would be interested in using these devices for crowdwork if HITs on these devices can be made sufficiently usable.

\section{Discussion}

In this section, we discuss what steps could be taken to begin addressing impediments to using non-workstation devices in crowdwork.
Critically, we expect that by addressing these issues, workers will feel more confident in adopting non-workstation devices into their workflow as appropriate, thereby increasing workers' flexibility, allowing them to complete work at the time, place, and manner of their choosing.
Doing so will not only benefits workers, but requesters as well, as workers with flexible work environments are more likely to produce high quality work~\cite{baltes1999flexible,neirotti2019designing}.

\subsection{Competing with Bots}

The most pressing issue for workers was being able to compete with bots for high-paying work.
To address this problem, they are relying on tooling that allows them to automatically find and accept work.
An obvious first step to supporting crowdwork management on non-workstation devices would be to support this tooling on those devices.
For smartphones and tablets, there is a desire to add support for all aspects of automatically finding and accepting work.
For smart speakers and smartwatches, support could be limited to receiving notifications and accepting particularly interesting work.

Still, such a solution is merely a band-aid and is insufficient to address such an important problem.
Critically, this problem applies not only to workers but also to requesters, as bots degrade the quality of answers received.
In the worst case, they ruin the ecological validity of user studies or defeat the purpose of using crowdwork to train ML/AI models.
Likely, crowdworking platforms are already devoting efforts to identifying and removing bots from their platforms.
As such, we look to participants' responses to identify other ways to help fight this problem that do not depend on bot detection but are additive to those efforts.

Our proposed solution revolves around removing the time-sensitivity around finding and accepting work.
Instead of assigning workers on a first-come-first-served basis, we propose having workers identify how much time they plan to spend crowdworking and then bid on the HITs that they find most compelling.
After a period of time, MTurk would gather all current bids, auto-assigning HITs to workers based on these bids and the worker's suitability to complete the HITs.
Workers would then have an extended period (16--24 hours) to complete the assigned HITs, freeing up the workers to complete the HITs at the time, location, and manner of their choosing.

Such a solution has clear benefits for workers.
First, it puts workers on equal footing with bots in terms of catching work, as there is no benefit to instantly submitting bids.
Second, workers can indicate how much work they want and rely on the platform to find sufficient work (or at least their fair share).
Third, the proposed system separates the process of managing HITs from completing those HITs, allowing users to use different devices for each activity.
Fourth, it removes the need to install specialized tooling.

However, such a system would require adjustments from workers and requesters.
For workers, it would require them to plan ahead regarding how much work they could do.
For requesters, it would introduce additional delay, as they would need to wait for at least a single bidding period and then the period for work to be completed.
Nonetheless, we expect this system's benefits to outweigh its costs.
Still, there is room for future research to further explore this idea, measuring how platforms, requesters, and workers would perceive it.
Research could also guide the selection of the length and frequency of bidding and work periods.

\subsection{Improving HIT Design}

Our participants thought that smartphones or tablets were not well suited for completing crowdwork.
Still, many participants wanted to see HIT design improved for the various devices: smartphones (35\%), tablets (26\%), smart speakers (7\%), and smartwatches (8\%).

The primary challenge identified by participants was the small screen sizes of non-workstation devices.
Luckily, responsive design is already a key tenant of web design, and the principles, methods, and tools that already exist for responsive web development can be applied to build HIT interfaces that will scale to any reasonable screen size (i.e., smartphones and tablets).
While requesters will need to exert more engineering effort, we feel that this is a cost worth paying, as it could substantially lessen the burden placed on workers when they try to complete these HITs.
Since requesters are receiving significant benefits from cheaply paid crowdwork, we believe they should do all they can to avoid burdening these users who rely on this work to earn a meager living (i.e., minimizing harm).
Moreover, responsive HIT design will increase workers' flexibility, allowing them to use non-workstation devices. This will also untether workers from the need to complete work at home, letting them complete work at the time, place, and manner of their choosing.
As prior research has shown~\cite{yin2018running}, this can potentially increase the quality of work done, directly benefiting requesters for their efforts.

The other major challenge identified by participants was needing to use a keyboard when entering text.
While it is impossible to add physical keyboards to these devices, we advocate for improved support for voice-based entry.
This could include (a) reminding users at the start of the survey that they are free to use voice transcription for their answers, (b) adding a button to text entry fields that activates on-device voice transcription, or (c) allowing users to record audio for their answers, which the researcher then transcribes.
We expect that supporting voice-based entry will not only allow crowdworkers to use their non-workstation devices, providing them with increased flexibility but could also improve the quality of their answers~\cite{wambsganss2022designing}.
In our experience, textual answers are terse, often providing little more than a few sentences worth of text.
In contrast, voice-based entry may encourage respondents to speak at greater length about topics of interest.
Moreover, by analyzing the audio, it may be possible to better understand the emotional context of the answer.
Due to these potential benefits, we strongly encourage requesters to adopt voice-based entry.

Finally, we believe there is room for creating HITs specifically tailored to non-workstation devices.
While prior research has explored using the audio recording capabilities of non-workstation devices~\cite{vashistha2017respeak,vashistha2018bspeak,hettiachchi2020hi,nebeling2016wearwrite}, its been 15 years~\cite{yan2009mcrowd} since there has been work exploring using these devices' other sensors.
Considering the emergence and mainstreaming of non-workstation devices globally in this time frame, we advocate for reexamining this area of research.
For example, using the health sensors integrated into smartwatches, biometrics could be collected as users complete tasks, something that has traditionally only been possible within lab settings.
Similarly, leveraging geolocation data could allow studies to examine how participants' locations (e.g., out in public, at home, at a store) impact their answers, opening new lines of inquiry.

\subsection{Smart Speakers and Smartwatches}
Our work shows a general lack of enthusiasm for using smart speakers or smartwatches to complete crowdwork. This stands in stark contrast to prior studies, which found some openness to using smart speakers for such tasks~\cite{nebeling2016wearwrite,hettiachchi2020hi,hettiachchi2020context}. We hypothesize that this discrepancy arises because, when asked hypothetical questions about using these devices for specific tasks, participants tend to focus on the novelty and respond with overly enthusiastic answers~\cite{ruoti2015authentication}. However, when considering how these devices would fit into their daily routines, as in our study, their enthusiasm diminishes and is replaced by skepticism.

Future work is needed to test this hypothesis. This could be achieved by creating various HITs that require the use of a smart speaker or smartwatch and measuring the participation rates. Additionally, participants could be surveyed after completing these HITs to gather feedback on their experiences with these devices. This approach would provide empirical evidence on whether participants are genuinely willing to use smart speakers or smartwatches for crowdwork.

Interestingly, while there is a lack of enthusiasm for completing tasks on these devices, a fifth of our participants expressed a desire for HIT management tools that could be used with smart speakers and smartwatches. They were particularly interested in features that would allow them to receive notifications about interesting tasks and accept those tasks while away from their workstations. This suggests that while participants may not see these devices as suitable for completing tasks, they do recognize their potential value in managing work more efficiently.

For platforms, this presents a strong motivation to explore the development of HIT management tools for smart speakers and smartwatches. By enabling workers to stay connected and manage tasks more flexibly, platforms could enhance worker satisfaction and increase overall engagement. This could also position the platform as an innovator in the field, attracting more workers who value flexibility and modern tools in their workflows.

As we are aware of no prior work investigating crowdwork management on these devices, we call upon the research community to pursue this direction. Developing and implementing management tools for smart speakers and smartwatches could significantly benefit workers, making crowdwork more accessible and integrated into their daily lives. For platforms, this could lead to higher retention rates, improved task completion, and a stronger competitive edge in the rapidly evolving gig economy.
\section{Conclusion}

In this paper, we examined crowdworkers' perceptions of using non-workstation devices---smartphones, tablets, smart speakers, and smartwatches---to complete and manage crowdwork.
Additionally, we explore workers' current practices and their desired practices in a world where crowdwork has been tailored to these non-workstation devices.
Our results show that crowdworkers find non-workstation devices unsuitable for crowdwork.
Regarding managing crowdwork, this unsuitability arises from the lack of crowdwork management tooling on non-workstation devices, tooling which workers find critical in their quest to outcompete bots in finding and accepting high-paying work.
Regarding completing crowdwork, participants face significant challenges completing current HITs on non-workstation devices, though they are somewhat open to using those devices if HIT design can be improved, particularly for smartphones and tablets.
Finally, workers are generally uninterested in using smart speakers or smartwatches to complete crowdwork, even if it is tailored to these devices; however, they are interested in using these devices to be notified of and accept interesting HITs.

We also propose several areas of future work to address the challenges identified in our work.
First, we propose a new paradigm for finding, accepting, and completing crowdwork that puts crowdworkers on equal footing with bots in these tasks.
Second, we describe how HIT design can be improved to increase the usability and utility of HIT completion across all non-workstation devices.
Third, we discuss how future studies can definitively answer whether participants are actually interested in using smart speakers and smartwatches in crowdwork.
We believe that this work has the potential to significantly improve workers' flexibility, allowing them to complete work in the time, place, and manner of their choosing.
Not only will this benefit workers, but we anticipate that increased flexibility will improve the quality of workers' responses~\cite{baltes1999flexible,neirotti2019designing}, providing tangible benefits to requesters (i.e., this community) and, by extension, the crowdwork platforms.

\bibliographystyle{unsrt}  
\bibliography{aaai24}

\begin{thebibliography}{10}

\bibitem{posch2019measuring}
Lisa Posch, Arnim Bleier, Clemens~M Lechner, Daniel Danner, Fabian Fl{\"o}ck, and Markus Strohmaier.
\newblock Measuring motivations of crowdworkers: The multidimensional crowdworker motivation scale.
\newblock {\em ACM Transactions on Social Computing}, 2(2):1--34, 2019.

\bibitem{yin2018running}
Ming Yin, Siddharth Suri, and Mary~L. Gray.
\newblock Running out of time: The impact and value of flexibility in on-demand crowdwork.
\newblock In {\em Proceedings of the 2018 CHI Conference on Human Factors in Computing Systems}, CHI '18, page 1–11, New York, NY, USA, 2018. Association for Computing Machinery.

\bibitem{baltes1999flexible}
Boris~B Baltes, Thomas~E Briggs, Joseph~W Huff, Julie~A Wright, and George~A Neuman.
\newblock Flexible and compressed workweek schedules: A meta-analysis of their effects on work-related criteria.
\newblock {\em Journal of applied psychology}, 84(4):496, 1999.

\bibitem{neirotti2019designing}
Paolo Neirotti, Elisabetta Raguseo, and Luca Gastaldi.
\newblock Designing flexible work practices for job satisfaction: the relation between job characteristics and work disaggregation in different types of work arrangements.
\newblock {\em New technology, work and employment}, 34(2):116--138, 2019.

\bibitem{dutta2022mobilizing}
Senjuti Dutta, Rhema Linder, Doug Lowe, Richard Rosenbalm, Anastasia Kuzminykh, and Alex~C Williams.
\newblock Mobilizing crowdwork: A systematic assessment of the mobile usability of hits.
\newblock In {\em CHI Conference on Human Factors in Computing Systems}, pages 1--20, 2022.

\bibitem{hettiachchi2020context}
Danula Hettiachchi, Senuri Wijenayake, Simo Hosio, Vassilis Kostakos, and Jorge Goncalves.
\newblock How context influences cross-device task acceptance in crowd work.
\newblock In {\em Proceedings of the AAAI Conference on Human Computation and Crowdsourcing}, volume 8(1), pages 53--62, 2020.

\bibitem{eagle2009txteagle}
Nathan Eagle.
\newblock txteagle: Mobile crowdsourcing.
\newblock In {\em International Conference on Internationalization, Design and Global Development}, pages 447--456. Springer, 2009.

\bibitem{narula2011mobileworks}
Prayag Narula, Philipp Gutheim, David Rolnitzky, Anand Kulkarni, and Bjoern Hartmann.
\newblock Mobileworks: A mobile crowdsourcing platform for workers at the bottom of the pyramid.
\newblock {\em Human Computation}, 11(11):45, 2011.

\bibitem{hettiachchi2020hi}
Danula Hettiachchi, Zhanna Sarsenbayeva, Fraser Allison, Niels van Berkel, Tilman Dingler, Gabriele Marini, Vassilis Kostakos, and Jorge Goncalves.
\newblock ``hi! i am the crowd tasker'' crowdsourcing through digital voice assistants.
\newblock In {\em Proceedings of the 2020 CHI Conference on Human Factors in Computing Systems}, pages 1--14, 2020.

\bibitem{nebeling2016wearwrite}
Michael Nebeling, Alexandra To, Anhong Guo, Adrian~A de~Freitas, Jaime Teevan, Steven~P Dow, and Jeffrey~P Bigham.
\newblock Wearwrite: Crowd-assisted writing from smartwatches.
\newblock In {\em Proceedings of the 2016 CHI conference on human factors in computing systems}, pages 3834--3846, 2016.

\bibitem{gupta2012mclerk}
Aakar Gupta, William Thies, Edward Cutrell, and Ravin Balakrishnan.
\newblock mclerk: enabling mobile crowdsourcing in developing regions.
\newblock In {\em Proceedings of the SIGCHI Conference on Human Factors in Computing Systems}, pages 1843--1852, 2012.

\bibitem{yan2009mcrowd}
Tingxin Yan, Matt Marzilli, Ryan Holmes, Deepak Ganesan, and Mark Corner.
\newblock mcrowd: a platform for mobile crowdsourcing.
\newblock In {\em Proceedings of the 7th ACM conference on embedded networked sensor systems}, pages 347--348, 2009.

\bibitem{kumar2014wallah}
Abhishek Kumar, Kuldeep Yadav, Suhas Dev, Shailesh Vaya, and G~Michael Youngblood.
\newblock Wallah: Design and evaluation of a task-centric mobile-based crowdsourcing platform.
\newblock In {\em Proceedings of the 11th international conference on mobile and ubiquitous systems: Computing, networking and services}, pages 188--197, 2014.

\bibitem{vashistha2017respeak}
Aditya Vashistha, Pooja Sethi, and Richard Anderson.
\newblock Respeak: A voice-based, crowd-powered speech transcription system.
\newblock In {\em Proceedings of the 2017 CHI conference on human factors in computing systems}, pages 1855--1866, 2017.

\bibitem{vashistha2018bspeak}
Aditya Vashistha, Pooja Sethi, and Richard Anderson.
\newblock Bspeak: An accessible voice-based crowdsourcing marketplace for low-income blind people.
\newblock In {\em Proceedings of the 2018 CHI Conference on Human Factors in Computing Systems}, pages 1--13, 2018.

\bibitem{williams2019perpetual}
Alex~C Williams, Gloria Mark, Kristy Milland, Edward Lank, and Edith Law.
\newblock The perpetual work life of crowdworkers: How tooling practices increase fragmentation in crowdwork.
\newblock {\em Proceedings of the ACM on Human-Computer Interaction}, 3(CSCW):1--28, 2019.

\bibitem{newlands2021crowdwork}
Gemma Newlands and Christoph Lutz.
\newblock Crowdwork and the mobile underclass: Barriers to participation in india and the united states.
\newblock {\em new media \& society}, 23(6):1341--1361, 2021.

\bibitem{boyatzis1998transforming}
Richard~E Boyatzis.
\newblock {\em Transforming qualitative information: Thematic analysis and code development}.
\newblock sage, 1998.

\bibitem{gioia2013seeking}
Dennis~A Gioia, Kevin~G Corley, and Aimee~L Hamilton.
\newblock Seeking qualitative rigor in inductive research: Notes on the gioia methodology.
\newblock {\em Organizational research methods}, 16(1):15--31, 2013.

\bibitem{wambsganss2022designing}
Thiemo Wambsganss, Naim Zierau, Matthias S{\"o}llner, Tanja K{\"a}ser, Kenneth~R Koedinger, and Jan~Marco Leimeister.
\newblock Designing conversational evaluation tools: A comparison of text and voice modalities to improve response quality in course evaluations.
\newblock {\em Proceedings of the ACM on Human-Computer Interaction}, 6(CSCW2):1--27, 2022.

\bibitem{ruoti2015authentication}
Scott Ruoti, Brent Roberts, and Kent Seamons.
\newblock Authentication melee: a usability analysis of seven web authentication systems.
\newblock In {\em Proceedings of the 24th International Conference on World Wide Web}. ACM, 2015.

\end{thebibliography}
\end{document}